\PassOptionsToPackage{dvipsnames}{xcolor}
\PassOptionsToPackage{hypertexnames=false,colorlinks=true,allcolors=BlueViolet}{hyperref}
\documentclass[aps,prl,reprint,superscriptaddress,nopreprintnumbers,nobibnotes,floatfix,longbibliography,nofootinbib,nolongbibliography]{revtex4-2}

\usepackage{newtxtext,newtxmath}
\usepackage{graphicx}
\usepackage[normalem]{ulem}
\usepackage{orcidlink}
\usepackage{titlesec}
\titleformat{\section}[runin]{\itshape}{}{0pt}{}[---]
\titlespacing{\section}{\parindent}{0pt}{0pt}
\titleformat{\subsection}[runin]{\upshape}{\thesubsection}{0pt}{}[:]
\titlespacing{\subsection}{\parindent}{0pt}{1em}

\usepackage{scalefnt}

\newcommand{\TNG}{\textsc{tng}}
\newcommand{\IllustrisTNG}{\textsc{IllustrisTNG}}
\newcommand{\TNGfifty}{\TNG\oldstylenums{50}}
\newcommand{\FIRE}{\textsc{Fire}-\oldstylenums{2}}

\newcommand{\Subfind}{Subfind}
\newcommand{\vphi}{\ensuremath{\bar{v}_\phi}}

\newcommand{\Erec}{\ensuremath{E_{\mathrm{\scriptscriptstyle NR}}}}

\newcommand{\xsec}{\ensuremath{\sigma_{\mathrm{\scriptscriptstyle SI}}}}
\newcommand{\mchi}{\ensuremath{m_\chi}}

\newcommand{\fRMS}{\ensuremath{f_{\mathrm{\scriptscriptstyle RMS}}}}
\newcommand{\kms}{\ensuremath{\text{km}~\mathrm{s}^{-1}}}
\newcommand{\skm}{\ensuremath{\text{km}^{-1}~\mathrm{s}}}
\newcommand{\vmin}{\ensuremath{v_\mathrm{min}}}
\newcommand{\eV}{\ensuremath{\text{e\kern-0.1emV}}}
\newcommand{\keV}{\ensuremath{\text{k\eV}}}

\newcommand{\GeV}{\ensuremath{\text{G\eV}}}
\newcommand{\TeV}{\ensuremath{\text{T\eV}}}
\newcommand{\GeVcm}{\ensuremath{\GeV~\mathrm{cm}^{-3}}}
\newcommand{\GeVc}{\ensuremath{\GeV}}
\newcommand{\osim}{\mathord{\sim}}

\newcommand{\Msun}{\ensuremath{\mathrm{M}_\odot}}
\newcommand{\Rsun}{\ensuremath{R_{\kern-0.1ex\odot}}}
\newcommand{\Rm}{\ensuremath{R_{\kern-0.2ex M}}}

\begin{document}

\title{Ubiquitous Corotation of Dark Matter Halos: Implications for Direct Detection}

\author{Dylan Folsom~\orcidlink{0000-0002-1544-1381}}
\email[Contact author: ]{dfolsom@princeton.edu}
\affiliation{Department of Physics, \href{https://ror.org/00hx57361}{Princeton University}, Princeton, NJ 08544, USA}

\author{Carlos Blanco~\orcidlink{0000-0001-8971-834X}}
\affiliation{Department of Physics, \href{https://ror.org/04p491231}{The Pennsylvania State University}, University Park, PA 16802, USA}
\affiliation{Department of Physics, \href{https://ror.org/00hx57361}{Princeton University}, Princeton, NJ 08544, USA}
\affiliation{\href{https://ror.org/05f0yaq80}{Stockholm University} and The Oskar Klein Centre for Cosmoparticle Physics, Alba Nova, 10691 Stockholm, Sweden}

\author{Mariangela Lisanti~\orcidlink{0000-0002-8495-8659}}
\affiliation{Department of Physics, \href{https://ror.org/00hx57361}{Princeton University}, Princeton, NJ 08544, USA}
\affiliation{Center for Computational Astrophysics, \href{https://ror.org/00sekdz59}{Flatiron Institute}, New York, NY 10010, USA}

\author{Mark Vogelsberger~\orcidlink{0000-0001-8593-7692}}
\affiliation{Department of Physics and Kavli Institute for Astrophysics and Space Research, \href{https://ror.org/042nb2s44}{Massachusetts Institute of Technology}, Cambridge, MA 02139, USA}

\date{\today}
\begin{abstract}
Cosmological simulations have recently begun to quantify the halo-to-halo variance in the phase-space distribution of dark matter around the Sun.
We use a sample of nearly one hundred Milky Way--like galaxies from the \TNGfifty{} simulation to determine what aspects of this variance control the predictions for dark matter direct detection.
Contrary to the isotropy assumed in the standard halo model, we find the dark matter median azimuthal velocity is nonzero and preferentially \emph{corotating}, i.e., in the direction of the baryonic disk's rotation, ranging from 6--70~\kms{}~(16th--84th percentile).
This corotation suppresses predicted scattering rates in laboratory experiments searching for dark matter lighter than 50~\GeVc{} and significantly affects the expected daily modulation amplitude for directional detectors.
In particular, this induces a 21\% uncertainty on the upper limit of the dark matter--nucleon interaction cross section at peak sensitivity for a typical isotropic ton-scale experiment.
This uncertainty is not irreducible, however: it is strongly correlated with the rotational velocity.
If studies of the Milky Way's formation history determine the rotation speed, this astrophysical uncertainty is reduced to 7\%.
\end{abstract}

\maketitle

\section{Introduction} 
\begin{figure}
    \includegraphics{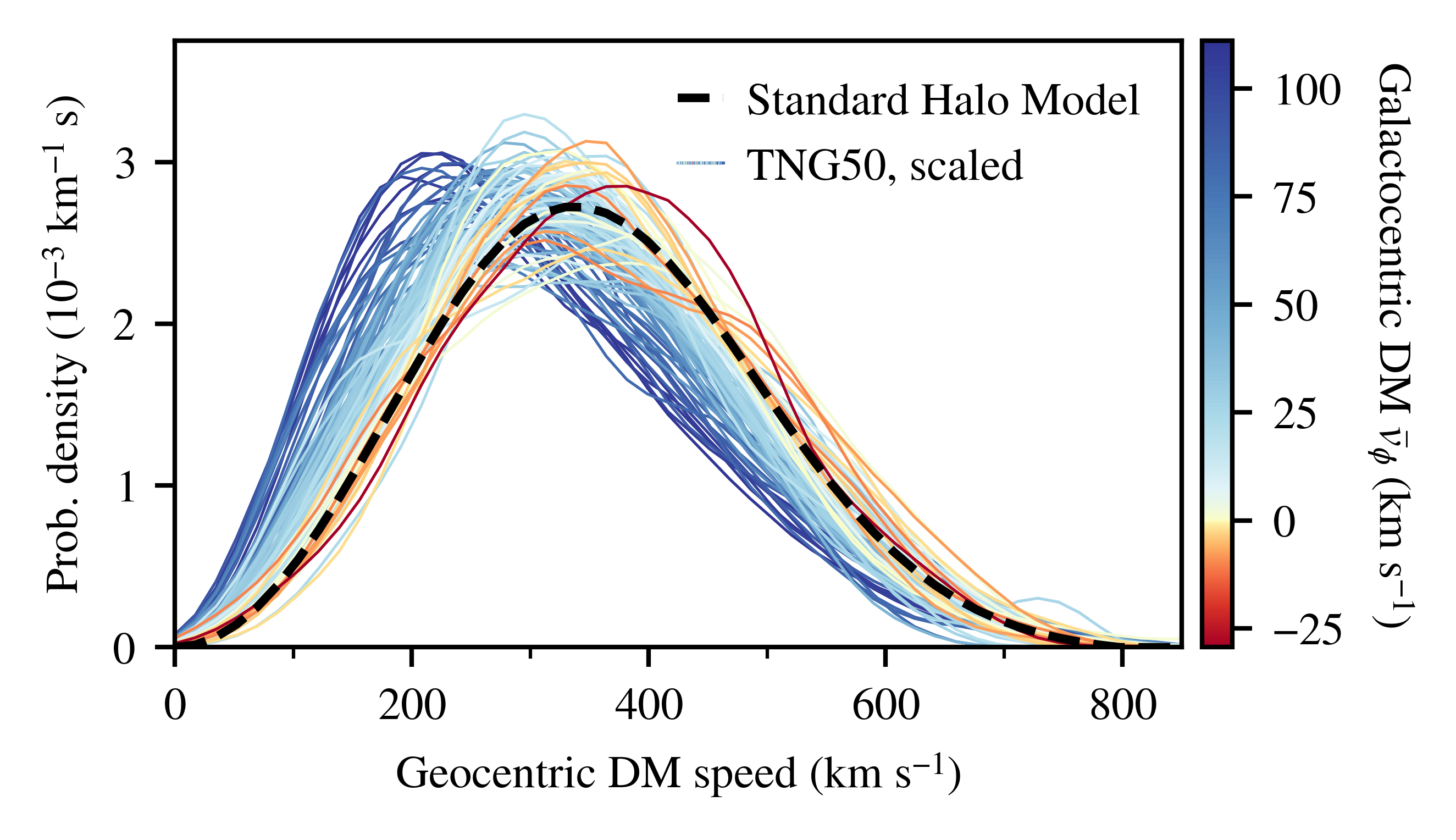}
    \caption{Geocentric speed distributions for DM in the solar neighborhood. The \TNGfifty{} halos are colored by the median DM azimuthal velocity, \vphi, after applying the scaling procedure of Ref.~\cite{Folsom:2025lly}. Halos with stronger corotation~(blue) exhibit slower geocentric speeds. Halos that rotate in the opposite direction from the baryonic disk~(red) have a headwind from this rotation, boosting the geocentric DM speeds relative to the SHM (dashed black).}
    \label{fig:fv_geo}
\end{figure}

The sensitivity of direct detection experiments depends on the dark matter~(DM) phase-space distribution in the Solar neighborhood, which is set by the Milky Way's~(MW's) formation history. Simulations of MW analogues have explored the local velocity distribution~\cite{Gnedin:2004cx, Wojtak:2005fe, Hansen:2005yj, Vogelsberger:2007ny, Bruch:2008rx, Read:2008fh, Vogelsberger:2008qb, Kuhlen:2009vh, Ling:2009eh, Purcell:2009yp, Read:2009iv, Schmidt:2009kz,Tissera:2009cm, Green:2010gw, Vogelsberger:2012sa, Kuhlen:2013tra, Butsky:2015pya, Zavala:2015neh, Bozorgnia:2016ogo, Kelso:2016qqj, Sloane:2016kyi, Bozorgnia:2017brl, Necib:2018igl, Artale:2019wee, Bozorgnia:2019mjk, Hryczuk:2020trm, Poole-McKenzie:2020dbo, Lawrence:2022niq, Sales:2022ich, Nunez-Castineyra:2023dui, Sheng:2023ecm, Staudt:2024tdq}, but only recently have they furnished the large statistical samples and resolution necessary to characterize and quantify its halo-to-halo variance~\cite{Folsom:2025lly,Lilie:2025wkr}. In this work, we study 98 MW-like halos from the \TNGfifty{} simulation~\cite{Nelson:2018uso,Pillepich:2019bmb,Nelson:2019jkf} to uncover the source of the uncertainty in direct detection experiments. We find that the simulated halos typically predict a nonzero angular momentum in the DM component aligned with the Galactic disk---i.e., \emph{corotation}. As \autoref{fig:fv_geo} shows, this corotation biases the local DM velocity distribution, modifying the geocentric speeds from the isotropic standard halo model~(SHM,~\cite{Wasserman:1986hh,Drukier:1986tm,Freese:1987wu}). Understanding the ramifications of this shift is crucial for interpreting DM detection results.

Laboratory direct detection experiments have constrained spin-independent DM--nucleon cross sections over \GeV{}--\TeV{} masses~\cite{LZ:2024zvo,PandaX:2024qfu, XENON:2025vwd,XENON:2019zpr,LUX:2016ggv,Lebedenko:2008gb, DEAP:2019yzn,DarkSide-50:2022qzh,CRESST:2019jnq,EDELWEISS:2016nzl, CDMS-II:2009ktb,SuperCDMS:2018gro,NEWS-G:2017pxg,NEWS-G:2024jms, Adhikari:2018ljm,PandaX:2022aac, PICO:2017tgi,PICO:2019vsc,PICASSO:2012ngj,CRESST:2022dtl,COUPP:2012jrk, XENON:2019gfn,PandaX:2022xqx,PandaX-II:2021nsg,CDEX:2022kcd, DarkSide:2022knj,DAMIC-M:2023gxo,DAMIC-M:2023hgj,DAMIC-M:2025luv, SENSEI:2020dpa,SENSEI:2024yyt,SuperCDMS:2020ymb,EDELWEISS:2020fxc, Essig:2017kqs}. Current experimental interpretations generally assume the SHM and do not consider the variance in the local DM distribution. Understanding the astrophysical uncertainty will become increasingly necessary as experimental exposures grow with the next generation of liquid-noble observatories~\cite{Aalbers:2022dzr}, when coherent neutrino--nucleus interactions become an irreducible background~\cite{LZ:2025igz,PandaX:2024muv,LZ:2024zvo}. To see through this ``neutrino fog,'' a variety of directional detection architectures have been proposed~\cite{Burgos:2007zz,DRIFT:2016utn,Leyton:2016nit,Shimada:2023vky,Santos:2013hpa,Vahsen:2011qx,Vahsen:2020pzb,Mayet:2013mpa,Sekiya:2004ma,Sekiya:2004fw,Belli:2020wfp,Blanco:2021hlm,Blanco:2026kda,Cook:2024cgm,Blanco:2025sgv,Blanco:2022pkt,DarkSide-20k:2023nla,Golovatiuk:2020krw,Rajendran:2017ynw,Ebadi:2022axg,Seidel:2022ofd,Caputo:2020sys,Cavoto:2019flp,Hochberg:2016ntt,PTOLEMY:2018jst,Hochberg:2025dom,Boyd:2022tcn,Drukier:2012hj,Matas:2025iqw}. These search for a daily modulation in the signal from the time-dependent DM phase-space distribution.

In this study, we find that the dominant factor setting the uncertainty in the local DM velocity distribution is the halo-to-halo spread in the azimuthal velocity. The corotation with the baryonic disk determines the relative velocity of the DM wind, suppressing the isotropic scattering rates for DM candidates below $\osim 50$~\GeV{} and reshaping the modulation signatures that are critical for directional discovery.

\begin{figure*}
    \includegraphics{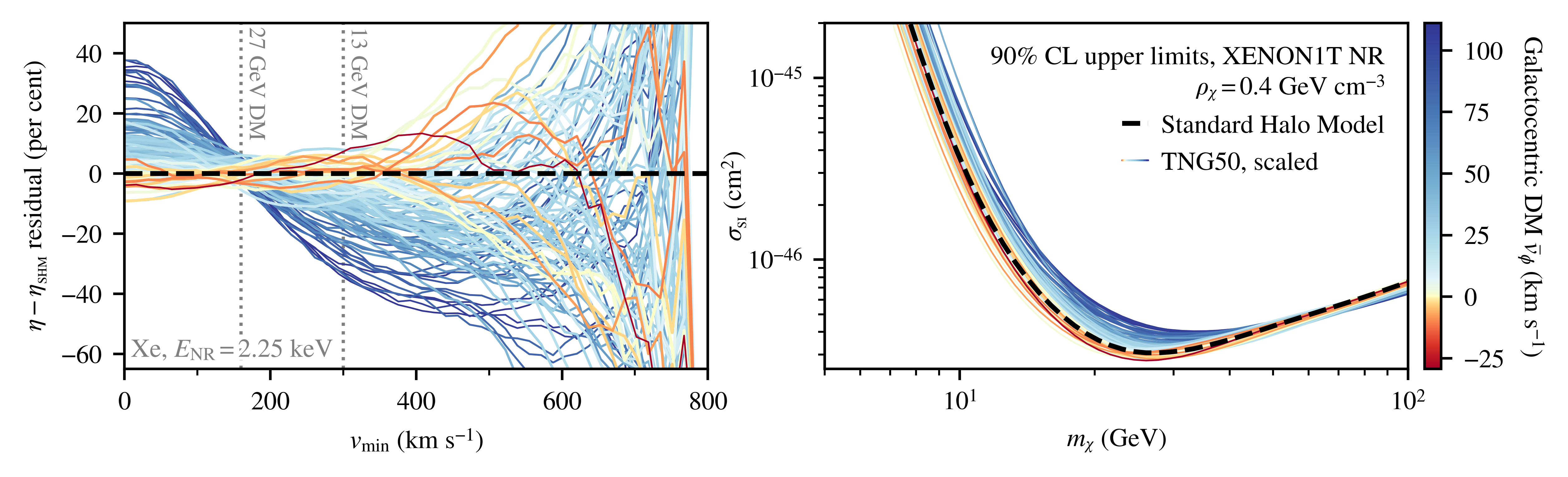}
    \caption{(Left) Residual of the halo integral $\eta$ (\autoref{eq:eta}) of the \TNGfifty{} MWs with respect to the SHM, colored according to the median azimuthal velocity \vphi{}. We show the minimum DM mass \mchi{} that could drive a 2.25~\keV{} nuclear recoil in xenon, corresponding to the threshold for the XENON1T experiment~\cite{XENON:2022zkh}, at $\vmin = 160$ and 300~\kms{} (dotted gray). For DM lighter than 27~\GeVc{}, corotating halos (in blue) lack DM with the speed required to cause such an event, reducing their $\eta$. For $\vmin\gtrsim 700$~\kms{}, $\eta$ probes the stochastically populated high-speed tail of the speed distribution, and the correlation with \vphi{} vanishes. (Right) 90\% upper limits on the DM--nucleon spin-independent cross section \xsec{}, computed for the XENON1T experiment. Below 50~\GeVc{}, limits from corotating halos are weaker than the SHM expectation.}
    \label{fig:xe_limits}
\end{figure*}

\section{Corotation in simulated halos}
We use a sample of 98~MW-like galaxies selected in Ref.~\cite{Folsom:2024add} from the \TNGfifty{} simulation~\cite{Nelson:2018uso,Pillepich:2019bmb,Nelson:2019jkf}, the highest resolution of the \IllustrisTNG{} magnetohydrodynamic suite, which consists of a $(51.7~\mathrm{Mpc})^3$ volume evolved from a redshift of 127 to redshift zero under a Planck cosmology~\cite{Planck:2015fie}. Halos are identified with the \Subfind{} algorithm~\cite{Springel:2000qu,Dolag:2008ar}, and the MW-like galaxies are selected from the \Subfind{} catalog such that (1) their stellar mass is in the range $(4\text{--}7)\times 10^{10}~\Msun$, (2) they are more than 500~kpc from the nearest larger halo, and (3) they are more than 1~Mpc from any halo with mass $\geq 10^{13}~\Msun$. 

Ref.~\cite{Folsom:2025lly} studied the DM speed distributions of these simulated MWs in the solar neighborhood and found that their enclosed mass at the solar position was often low compared to our Galaxy. To address this, they introduced a coordinate scaling procedure that sets the monopole term of each galaxy's gravitational potential to the expected MW value. This adjustment yields DM speed distributions that recover the local standard of rest speed, putting all the simulated halos on equal footing. In this work, we present all results for the scaled halos.

The analysis focuses on the DM in the solar neighborhood, viz. at heights $|z| \leq 1$~kpc from the disk plane and at cylindrical radii within one kiloparsec of the Sun's radius, $\Rsun=8.3$~kpc~\cite{Abuter:2021yys,GRAVITY:2024tth}. The \TNGfifty{} coordinate system is oriented with the $z$-axis parallel to the angular momentum of stars and star-forming gas within twice the stellar half-mass radius~\cite{2024MNRAS.535.1721P}, and the $+\hat{\phi}$ direction points with this rotation. Most halos have velocity distributions with nonzero median azimuthal velocity \vphi{}, with rotation speeds of $\vphi = 31_{-24}^{+39}$~\kms{} (16th, 50th, and 84th percentiles). The corotation can also be quantified in terms of the angular momentum of the DM in the solar neighborhood. At the present day, this is preferentially aligned with the galaxy's baryonic component, with a $5^{+29}_{-4}$~degree offset from the $z$-axis. The deviation from isotropy is only significant in the azimuthal direction: the corresponding radial and vertical median velocities are $\bar{v}_r = 0_{-3}^{+4}$~\kms{} and $\bar{v}_z = 1_{-6}^{+5}$~\kms{}, respectively. This velocity anisotropy does not meaningfully correspond to an anisotropic mass distribution: the ratio of major to minor principal inertial moments for DM in the spherical shell of radius $\Rsun{}\pm 1$~kpc is $0.89_{-0.04}^{+0.04}$, consistent with expectations for the inner halo~\cite{Chua:2021oqe}.

We boost the DM velocities into the Earth's frame as it was on March~9,~2000,\footnote{This frame has a Galactocentric velocity of 251~\kms{} azimuthally with the disk, 40~\kms{} toward the Galactic Center, and 12~\kms{} antiparallel to the disk's angular momentum (i.e., toward the north galactic pole)~\cite{Freese:2012xd,wimprates}.} when the geocentric speed distributions are typical of the annual average~\cite{Baxter:2021pqo}. \autoref{fig:fv_geo} shows the resulting DM speed distributions for each halo, with the SHM prediction in dashed black. Following the conventions of Ref.~\cite{Baxter:2021pqo}, the SHM is a Maxwell--Boltzmann distribution with mode $v_0 = 238$~\kms{} truncated at the escape speed 544~\kms{}.

The net effect of corotation is a slowing of the geocentric DM speeds. In \autoref{fig:fv_geo}, the speed distributions are colored according to \vphi{}, with strong corotation shown in blue, non-rotating halos in yellow, and retrograde rotation shown in red. The Earth's motion about the Galactic center boosts the DM speeds to a median of 351~\kms{} in the isotropic SHM. In the simulated sample, however, corotation diminishes the effect of this boost, and the median geocentric DM speed is $327^{+23}_{-25}$~\kms{}.

\section{Effects on DM direct detection} 
The slower geocentric speeds of the corotating DM modify the elastic DM--nucleon recoil spectrum. This can be parameterized by the halo integral $\eta$. For a DM particle of mass \mchi{} scattering with recoil energy \Erec{}, the differential scattering rate is
\begin{equation}\label{eq:eta}
    \frac{\mathrm{d}R}{\mathrm{d}\Erec} \propto \eta(\vmin)  = \int_{\vmin}^\infty \frac{f(\mathbf{v}_\mathrm{geo})}{|\mathbf{v}_\mathrm{geo}|}\;\mathrm{d}^3\mathbf{v}_\mathrm{geo} \,,
\end{equation}
where $f(\mathbf{v}_\mathrm{geo})$ is the geocentric DM velocity distribution and \vmin{} is the slowest DM speed that can drive recoils of energy \Erec{}~\cite{Lin:2019uvt}. The halo integral describes the expected scattering rate purely at the level of the DM kinematics, without the details of the detector apparatus. 

\begin{figure*}
    \includegraphics{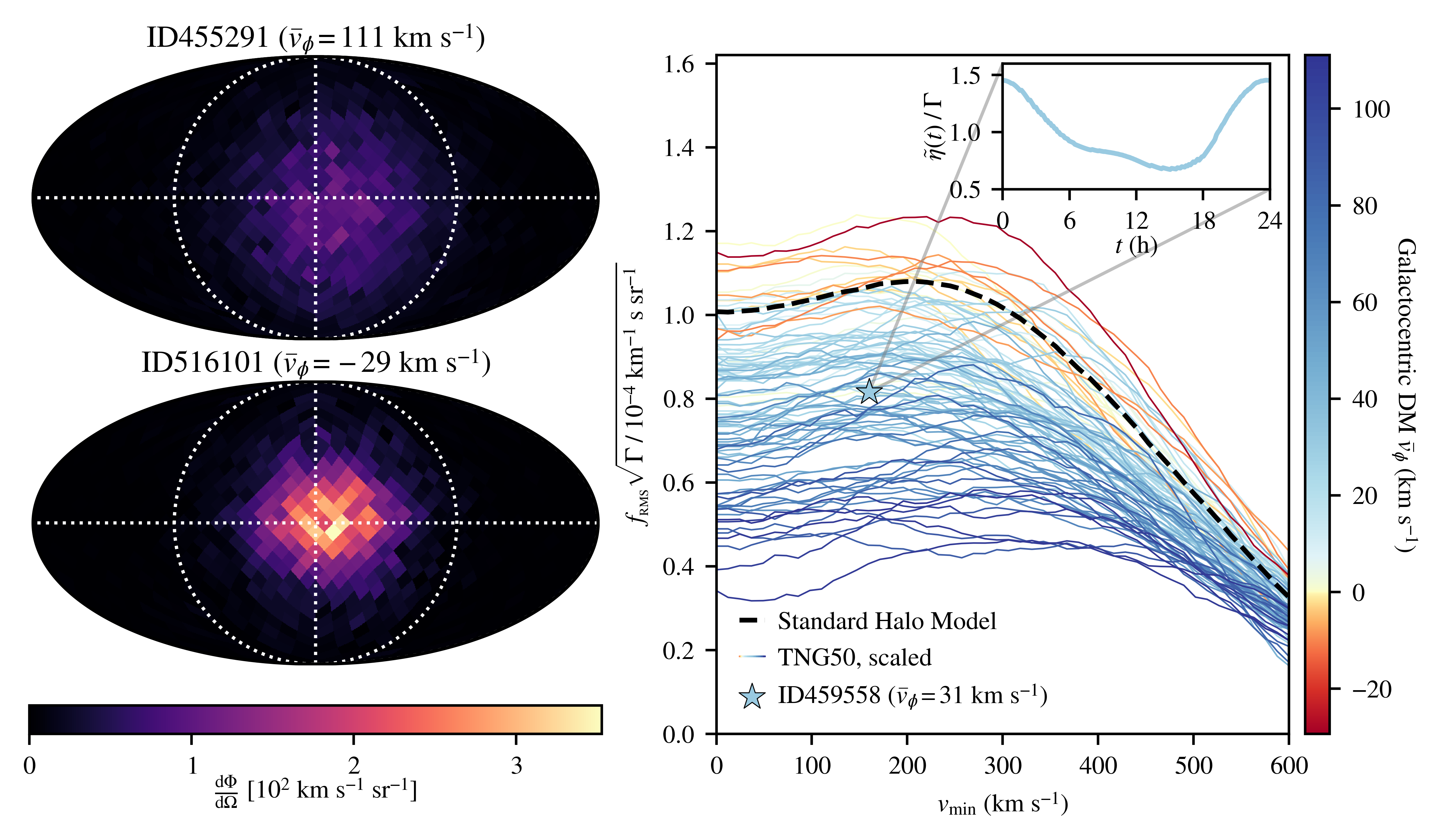}
    \caption{(Left) Distribution of geocentric DM flux projected onto the sky. The $+\hat{\phi}$ direction is in the center, with dotted white lines marking $90^\circ$ of longitude and the Galactic equator. The top (bottom) panel shows the halo with the greatest (lowest) \vphi{}. In both cases, the DM wind blows from the $+\hat{\phi}$ direction, but the flux is reduced and spread over a larger angular area in the corotating halo. (Right) The figure of merit $\fRMS{} \sqrt{\Gamma}$, where \fRMS{} is the RMS daily modulation of the differential halo integral $\tilde{\eta}$, and $\Gamma$ is the mean of $\tilde{\eta}$. This quantity is a proxy for the sensitivity of a directional detector~\cite{Blanco:2026kda}. It is generically suppressed in the corotating halos (blue) relative to the SHM (dashed black) because the diffuse signal reduces \fRMS{}. (Right, inset) An example of daily modulation in $\tilde{\eta}/\Gamma$ for the median-\vphi{} halo, evaluated at $\vmin{}=160$~\kms{}. }
    \label{fig:daily_mod}
\end{figure*}

The impact of corotation on the halo integral is shown in the left panel of \autoref{fig:xe_limits}, where the residual of each halo's $\eta$ is shown with respect to the SHM prediction, colored according to its \vphi{} as in \autoref{fig:fv_geo}. When $\vmin{} \lesssim 160$~\kms{}, the integral of \autoref{eq:eta} is taken over a large portion of velocity space and $\eta$ is approximately the average inverse geocentric speed. Because corotation slows the geocentric speeds, the highest-\vphi{} halos (in blue) have $\eta \sim 40\%$ larger than the SHM prediction. At higher \vmin{},  the $\eta$ of the corotating halos is lower because of the reduced population of high-speed DM relative to the SHM (see \autoref{fig:fv_geo}). This suppresses the differential scattering rates for the corotating halos as not enough particles move fast enough to induce nuclear recoils of the requisite energy. The highest-speed tails ($v_\mathrm{geo} \gtrsim 700$~\kms{}) are stochastically populated and have no correlation with \vphi{}. For a quantification of the correlation at all \vmin{}, see \hyperref[app:figs]{Appendix A}. 

The right panel of \autoref{fig:xe_limits} applies the recoil spectra derived from $\eta$ to a mock analysis. We compute the exact recoil spectra using the \texttt{wimprates} package~\cite{wimprates} and process them with an approximate XENON1T search pipeline from Ref.~\cite{XENON:2022zkh}. This pipeline emulates the analysis of 1~ton-yr of exposure from the XENON Collaboration~\cite{XENON:2019ykp,XENON:2018voc}, producing the value of the spin-independent DM--nucleon cross section \xsec{} for which the predicted signal would exceed the modeled detector background (for recoil energies of 2.25--58~\keV{}) at 90\% confidence level. \autoref{fig:xe_limits} shows this constraint on \xsec{} as a function of the DM mass \mchi{}, assuming a 0.4~\GeVcm{} DM density $\rho_\chi$, consistent with measurements at the Solar position~\cite{deSalas:2020hbh}.

When $\mchi \gtrsim 80$~\GeVc{}, most DM particles have sufficient kinetic energy to produce a detectable nuclear recoil, and the recoil rate is proportional to the average inverse DM speed. The corotating halos therefore have the highest rate of nuclear recoils and set $\lesssim 15\%$ stronger limits on \xsec{}. At and below peak sensitivity, $\mchi = 27$~\GeVc{}, only the fastest DM particles produce recoils in the signal region, and the corotating halos have $\gtrsim40\%$ weaker limits on \xsec{}. The constraints set at the lightest DM masses ($\mchi \lesssim 10$~\GeVc{}) are sensitive to the high-speed tail of the DM speed distribution, with the weakest limits 150\% weaker than the SHM prediction and no strong correlation between the limit and \vphi{}.

At peak sensitivity, the \xsec{} limits correlate strongly with \vphi{}. Because of this, the astrophysical uncertainty in the limit is not irreducible, and knowledge of \vphi{} reduces the uncertainty further. Consider the \TNGfifty{} halos' distribution of \vphi{} and \xsec{} upper limit as a two-dimensional Gaussian. The marginal distribution of \xsec{}, with no knowledge of \vphi{}, has 21\% uncertainty (i.e., a factor of $1.21\times$ from 16th--84th percentiles). However, because of the strong correlation between \vphi{} and the limit, the Gaussian has high covariance. The intrinsic uncertainty on the limit, given by the conditional distribution of \xsec{} at fixed \vphi{}, has a much smaller uncertainty: only 7\%. This reducibility is most significant at peak sensitivity, where the correlation with \vphi{} is strongest and detailed knowledge of the MW's DM distribution is most important.

While the above limits are derived assuming isotropic scattering, the anisotropy of corotating halos has distinct implications for directional DM detectors. The left panels of \autoref{fig:daily_mod} show the distribution of geocentric DM flux on the sky in a Mollweide projection. The top and bottom panels contrast the halos with the largest and smallest azimuthal velocities ($\vphi{} = 111$ and $-29$~\kms{}, respectively) to highlight the effects of corotation. The latter has a higher, more collimated total flux, with 68\% contained within a $43^\circ$ radius. In contrast, the highly corotating halo has a lower, more diffuse total flux spread over a $61^\circ$ radius. Across all simulated halos, this 68\% containment radius is typically $(51_{-4}^{+6}{})^\circ$, wider than the $47^\circ$ predicted by the SHM.

Directional DM detectors are sensitive to this on-sky distribution.\footnote{If the directional detector measures DM--electron scattering rather than DM--nucleon scattering, it modifies the relationship between \vmin{} and \mchi{}.} Consider a mock idealized experiment with a preferential axis for DM scattering. Defining $\vartheta$ as the angle between the DM-imparted momentum and the preferred axis, the probability of DM scatterings in such a material is proportional to $\cos^2\vartheta$ to leading nontrivial order~\cite{Lillard:2026jdo}. We take this expression as the point-spread function for our idealized detector---see \hyperref[app:detector]{Appendix B}. The orientation of the detector is chosen such that the preferred axis initially aligns with the Earth's Galactocentric velocity vector, with the DM wind optimally scattering within the detector. As the Earth rotates, however, the orientation of the detector will change, yielding a modulating signal.

To determine the signal modulation, consider the on-sky distribution of the halo integral, i.e., integrating \autoref{eq:eta} over the velocity magnitude and leaving the angular components. Convolving this with the detector's point-spread function yields $\tilde{\eta} = \eta\ast\cos^2\vartheta$. This is a function of the sky location---see \hyperref[app:figs]{Appendix A}---and is maximal in the directions parallel and anti-parallel to the Earth's Galactocentric velocity. We sample the $\tilde{\eta}$ distributions along the aforementioned trajectory, beginning in the direction of the Earth's velocity and tracing out a circle about the celestial north pole, emulating what a detector on Earth would see over the course of a day. We compute the root-mean-square (RMS) amplitude of $\tilde{\eta}$ along this trajectory,
\begin{equation}\label{eq:fRMS}
    \fRMS{} = \sqrt{\int_0^{T_{24}}\frac{\mathrm{d}t}{T_{24}}\left(\frac{\tilde{\eta}(t)}{\Gamma} - 1\right)^2}
\end{equation}
for $T_{24}=24$~h the modulation period and $\Gamma$ the daily average $\tilde{\eta}(t)$. The inset panel of \autoref{fig:daily_mod} shows an example $\tilde{\eta}(t)$ for the median-\vphi{} halo at $\vmin=160$~\kms{}, with $\Gamma = 10^{-3}~\skm~\mathrm{sr}^{-1}$ and $\fRMS = 26\%$.

While the sensitivity of a typical counting experiment scales with the square root of the exposure, a highly modulated signal is much easier to separate from a steady background. In such cases, the sensitivity scales proportionally with the signal's RMS modulation amplitude~\cite{Blanco:2026kda}. Therefore, we take  $\fRMS \sqrt{\Gamma}$ as the figure of merit approximating the relative sensitivity of a directional detector. This is not exactly the same as the expression of Ref.~\cite{Blanco:2026kda}---namely, our $\Gamma$ and \fRMS{} are based on $\eta$ rather than the full scattering rate---but it is comparable: $\tilde{\eta}$ is proportional to the expected signal rate at fixed recoil energy, so its RMS modulation and daily average should give a sense for the statistics of the total scattering rate. The figure of merit is shown in the right panel of \autoref{fig:daily_mod}, where the simulated halos are colored according to their \vphi{} and the SHM prediction is shown in dashed black.

In general, the figure of merit is reduced relative to the SHM prediction for the corotating halos. Above $\vmin = 160$~\kms{}, this is in part because of the suppression of $\eta$ (see \autoref{fig:xe_limits}), but the primary difference is in the angular size of the signal. Except at the highest \vmin{}, $\tilde{\eta}$ has support across a large swath of the sky, reducing the modulation in the signal (see \hyperref[app:figs]{Appendix A}). At $\vmin = 0$, $\fRMS{} = 23_{-8}^{+5}\%$ for the \TNGfifty{} halos, compared to the 29\% SHM prediction. There is a similar reduction in \fRMS{} at higher \vmin{}, which continues to suppress the expected sensitivity for corotating halos.. As with the isotropic case, however, this is strongly correlated with \vphi{}, such that a determination of this value for the MW would reduce the projected uncertainty in discovery significance. 

\section{Conclusions}
This work investigated the astrophysical uncertainties on DM direct detection bounds originating from the local velocity distribution of the MW halo. Using a sample of 98 MW--like galaxies from the \TNGfifty{} simulation, we found that the primary epistemic uncertainty in these bounds results from a 6--70~\kms{} median azimuthal velocity \vphi{} (16th--84th percentile) in the local velocity distribution.  Corotation, i.e., DM angular momentum parallel with that of the baryonic disk, is typical across the simulated sample, with 91\% of halos exhibiting $\vphi > 0$. Isotropy in the velocity distribution---a foundational assumption of the SHM---is rare. 

Nonzero \vphi{} has also been observed in the \FIRE{} Latte suite~\cite{Zhang:2026qnl}, and the distribution of \vphi{} values matches the \TNGfifty{} sample with a Kolmogorov--Smirnov $p$-value of 99\%.  While only 6 \FIRE{} halos are included, these results suggest that the corotation of MW-like halos persists across different sub-grid models. These preliminary results should be further validated with larger samples of MW-like halos, simulated at higher resolutions and with different sub-grid assumptions. We defer a detailed study of the mechanisms that set this corotation, but note that it is present in the \TNGfifty{} halos even at early times, while they are accumulating most of their present-day solar neighborhood material, see \hyperref[app:figs]{Appendix A}.  

The variations in DM flux yield a spread of 21\% (16th--84th percentile) in the peak-sensitivity upper limits on the DM--nucleon spin-independent cross section in a fiducial ton-scale liquid-noble detector, biased toward weaker limits than the SHM prediction. Further, the DM flux is spread over a larger region of the sky in the corotating halos than expected from the SHM. As a result, the amplitude of daily modulation in a directionally-sensitive detector is suppressed, reducing the expected sensitivity for such searches by $\lesssim 70\%$.

These effects correlate strongly with \vphi{}, indicating that the halo-to-halo variance that drives the astrophysical uncertainty in direct detection is not irreducible. Indeed, the dominant source of velocity distribution--induced uncertainty is a definitively measureable quantity: studies of the MW, e.g., Refs.~\cite{Necib:2018iwb,Necib:2018igl,Zhu:2024sve,Shpigel:2025ulk,Zhang:2026qnl}, have already begun to probe details of the local DM velocity distribution. Determining the extent of the MW's corotation---and therefore the expected DM signal spectrum---can reduce the astrophysical uncertainty to the percent level for the next generation of DM searches, a matter of increasing importance as we enter the neutrino fog. 

\vspace{\baselineskip}
\section{Acknowledgments}
We thank Lina Necib, Tal Shpigel, and Xiuyuan Zhang for useful conversations and their data. M.L. and D.F. are supported by the Department of Energy~(DOE) under Award No. DE-SC0007968. M.L. is also supported by the Simons Investigator in Physics Award. D.F. is additionally supported by the Joseph H. Taylor Graduate Student Fellowship. The computations in this paper were run on the FASRC cluster supported by the FAS Division of Science Research Computing Group at Harvard University, and we thank Lars Hernquist for his continued sponsorship in this regard. The \IllustrisTNG{} simulations were undertaken with compute time awarded by the Gauss Centre for Supercomputing~(GCS) under GCS Large-Scale Projects GCS-ILLU and GCS-DWAR on the GCS share of the supercomputer Hazel Hen at the High Performance Computing Center Stuttgart~(HLRS), as well as on the machines of the Max Planck Computing and Data Facility~(MPCDF) in Garching, Germany.

This report was prepared as an account of work sponsored by an agency of the United States Government. Neither the United States Government nor any agency thereof, nor any of their employees, makes any warranty, express or implied, or assumes any legal liability or responsibility for the accuracy, completeness, or usefulness of any information, apparatus, product, or process disclosed, or represents that its use would not infringe privately owned rights. Reference herein to any specific commercial product, process, or service by trade name, trademark, manufacturer, or otherwise does not necessarily constitute or imply its endorsement, recommendation, or favoring by the United States Government or any agency thereof. The views and opinions of authors expressed herein do not necessarily state or reflect those of the United States Government or any agency thereof.


\bibliography{main}
\onecolumngrid

\clearpage
\begin{center}
\large\textbf{End Matter}
\phantomsection
\label{app:figs}
\end{center}
\setcounter{section}{0}
\renewcommand{\theequation}{(\Alph{section}\arabic{equation})}
\twocolumngrid
\section{Appendix A:\enspace{}Supplemental figures}
\begin{figure}
    \includegraphics{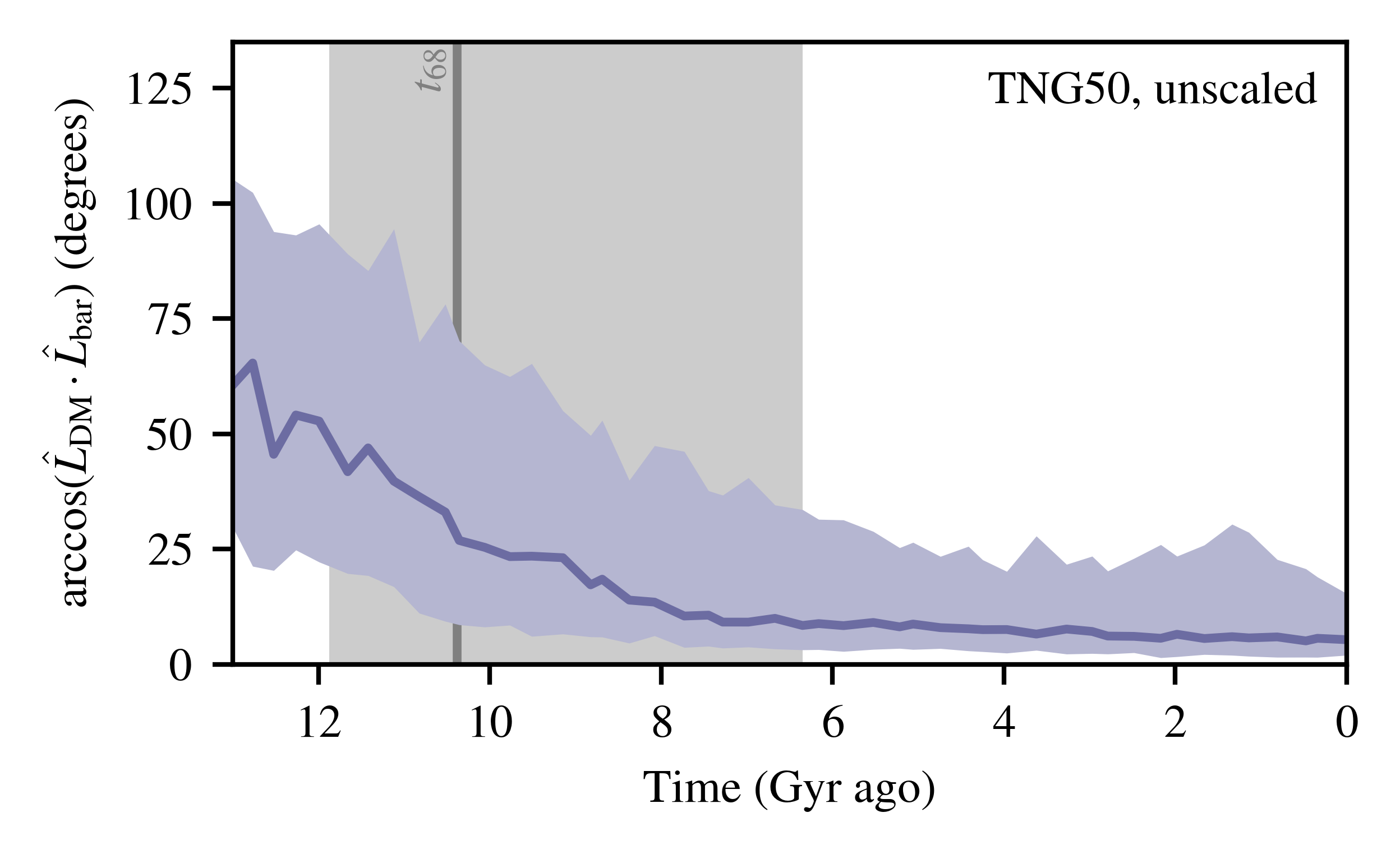}
    \caption{The angle between the angular momentum of the DM and the angular momentum of the baryonic matter across cosmic time. The purple shaded region encompasses the 16th--84th percentile range across the simulated sample, with a solid line denoting the median. The gray shaded region denotes the 16th--84th percentile range of $t_{68}$, the time at which the halos accumulate 68\% of the DM that ends up in the solar neighborhood, with a vertical line denoting the median value. At early times, the angular momentum of the DM and baryons is not necessarily aligned, but by the time most of the mass has assembled, the angular momenta tend to be aligned.}
    \label{fig:assembly}
\end{figure}

Here, we include figures that may be of interest to the reader. 

\autoref{fig:assembly} highlights the alignment between the DM and baryonic angular momenta even at early times.   In detail, it shows the angle between the DM and baryonic angular momentum in purple, with a band spanning the 16th--84th percentiles and a solid line denoting the median. The angular momenta are computed---without the scaling procedure of Ref.~\cite{Folsom:2025lly}---within twice the instantaneous stellar half-mass radius. This is shown as a function of lookback time through the simulation. Very early on, the angular momenta are not particularly aligned, with $(65^{+37}_{-44})^\circ$ between them, but within a few Gyr these momenta come into alignment, such that at the present day they are offset by only $(5^{+29}_{-4})^\circ$. This period of alignment coincides with when the halos assemble most of their mass. For each halo, we record the particle IDs of the DM that comprises the present-day solar neighborhood. At each snapshot, we check what fraction of that DM is bound to the halo. The quantity $t_{68}$ is the time at which 68\% of the material is bound, and gives a sense for when the halos formed. In the figure, the 16th--84th percentile range of $t_{68}$ is shown in gray, with a solid line denoting the median. As these halos assemble, they accumulate material through the accretion of satellite galaxies, and the DM and baryonic matter from these galaxies is accreted with similar angular momenta. This co-accretion process may be the source of the present-day corotation. 

\begin{figure}
    \includegraphics{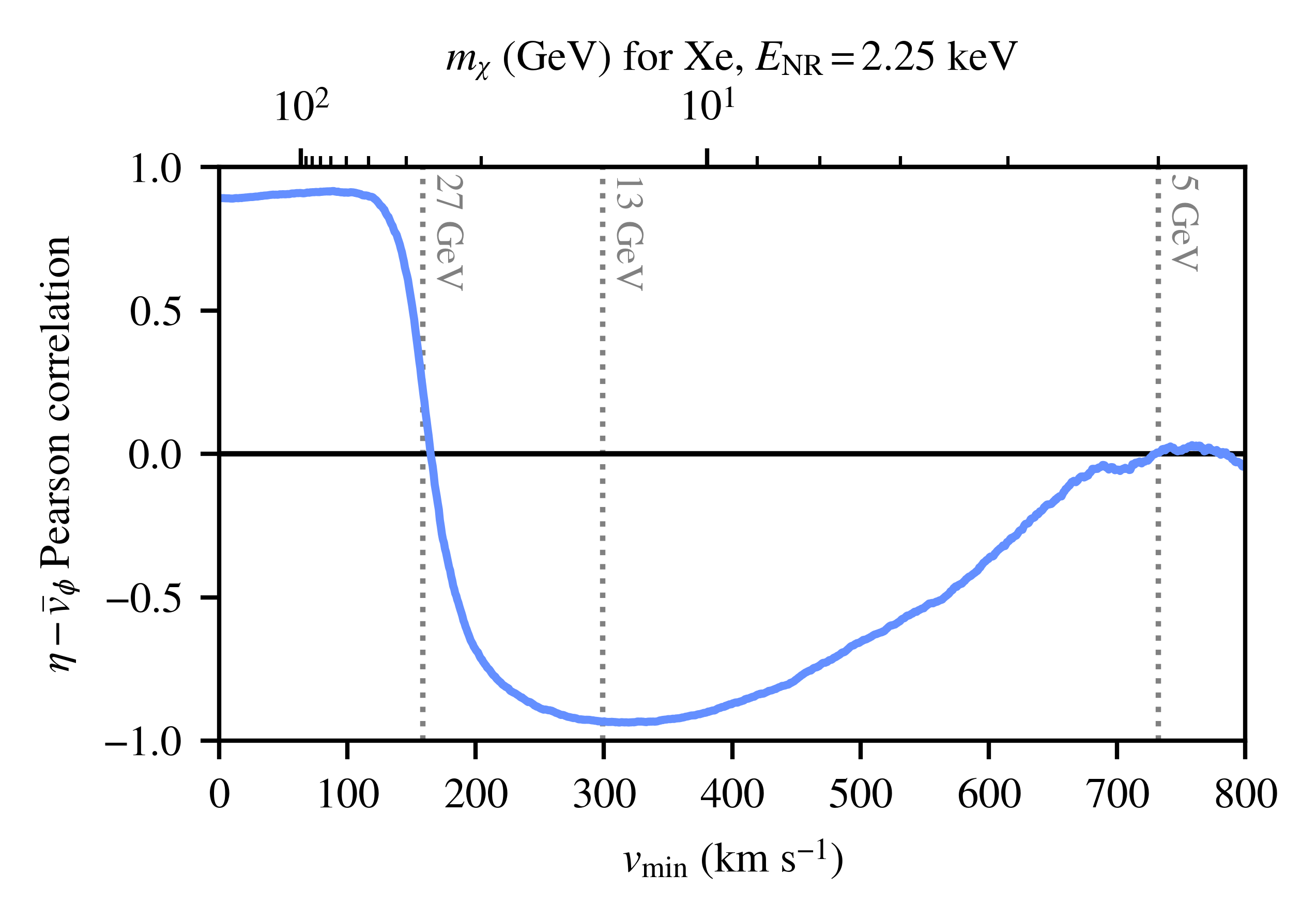}
    \caption{Correlation between $\eta$ and \vphi{}. To guide the eye, we show on the top axis (and with dotted gray lines) the minimum DM mass \mchi{} that, at a speed \vmin{}, can drive a recoil of 2.25~\keV{} in xenon, corresponding to the threshold energy of the XENON1T experiment~\cite{XENON:2022zkh}. \vphi{} is strongly correlated with $\eta$ for low \vmin{}, but above $\osim160$~\kms{} becomes anticorrelated. At very large \vmin{}, the halo integral is performed over only the stochastically-populated high-velocity tail and the correlation with \vphi{} approaches zero.}
    \label{fig:eta-vphi_correlation}
\end{figure}
\begin{figure*}
    \includegraphics[width=\textwidth]{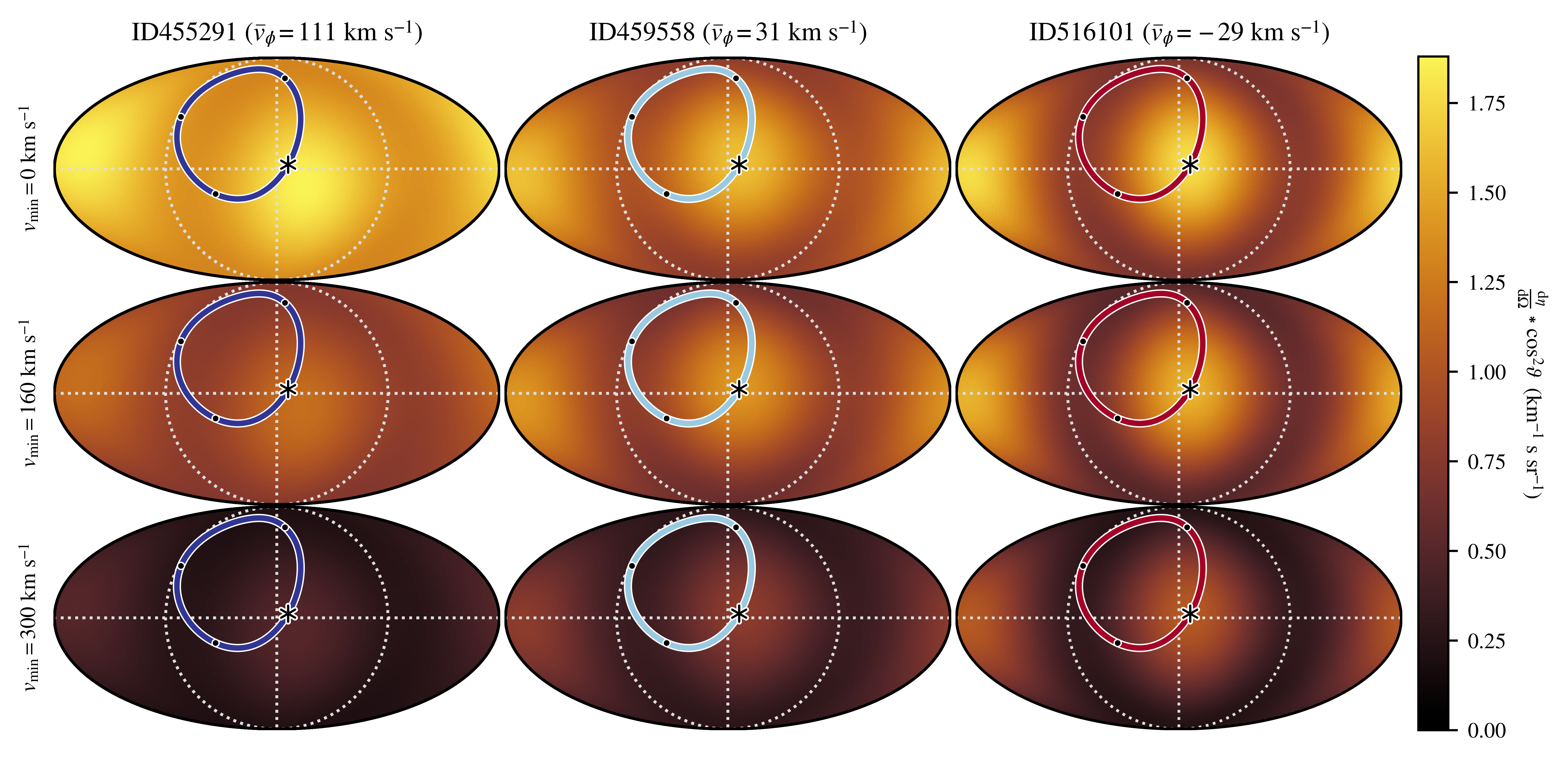}
    \caption{Mollweide projections of $\tilde{\eta}$, the halo integral convolved with a $\cos^2\vartheta$ kernel to emulate the signal seen by an idealized detector. The $+\hat{\phi}$ direction is in the center of each panel, and dotted white lines mark $90^\circ$ of longitude and the Galactic equator. Three halos are shown: the halo with the greatest, median, and least \vphi{} in the left, center, and right columns, respectively; their $\tilde{\eta}$ is shown at $\vmin = 0$, 160, and 300~\kms{}. The total $\tilde{\eta}$ decreases with increasing \vmin{}, and this decline is sharpest in strongly corotating halos (see \autoref{fig:xe_limits}, left panel). The colored lines denote the trajectory used to compute \fRMS{}: a circle centered at the north celestial pole, intersecting the Earth's Galactocentric velocity vector (marked with an asterisk). Further points are drawn every six hours across the circle to guide the eye.}
    \label{fig:eta_grid}
\end{figure*}
\begin{figure}
    \includegraphics{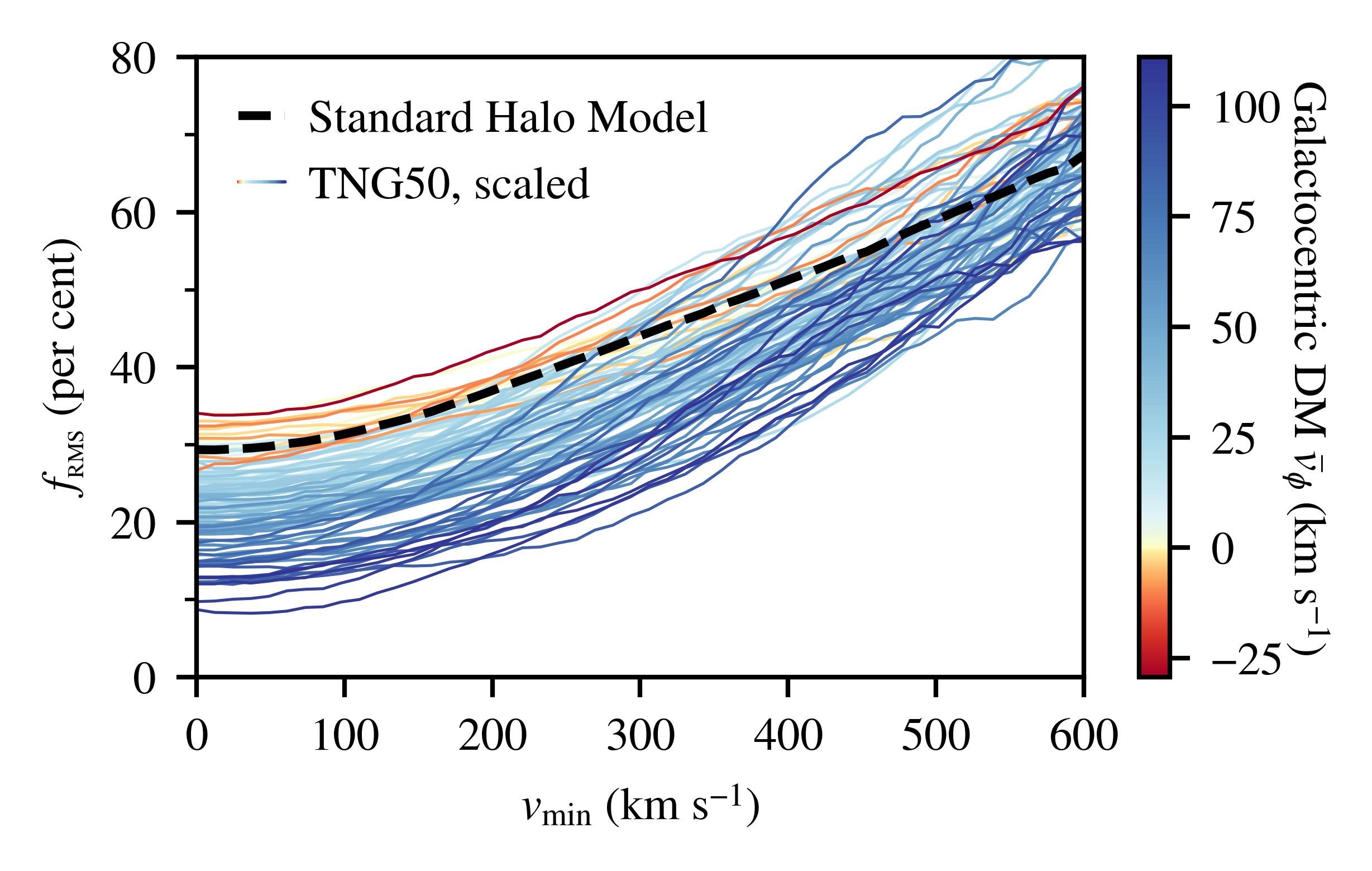}
    \caption{On-sky modulation amplitude of $\tilde{\eta}$, \fRMS{}. This is suppressed in high-\vphi{} halos and at low \vmin{}, as in these cases DM from many directions contributes to $\tilde{\eta}$. This reduces the expected sensitivity of DM searches in halos with strong corotation (see \autoref{fig:daily_mod}).}
    \label{fig:fRMS}
\end{figure}

As a companion to \autoref{fig:xe_limits}, \autoref{fig:eta-vphi_correlation} shows the correlation between $\eta$ and \vphi{} across a range of \vmin{} for an isotropic detector.   Specifically, it plots the Pearson $r$ between $\eta$ and \vphi{} for a range of \vmin{}. For reference, the upper axis of the figure shows the corresponding minimum DM mass \mchi{} that can drive a 2.25~\keV{} recoil in xenon when moving at a speed \vmin{}, and we highlight a few particular DM masses with dotted gray lines. At low \vmin{}, below $\osim160$~\kms{}, only heavy DM is able to excite such a transition, but the slower geocentric speeds of the corotating halos lead to a higher value of $\eta$, yielding a strong correlation between $\eta$ and \vphi{}. This asymptotes to $r^2=80\%$ as \vmin{} approaches zero. The quantities are maximally anticorrelated at $\vmin{} = 313$~\kms{}, with an $r^2 = 88\%$. At very large \vmin{}, the halo integral is performed over only the highest-speed tail, which is stochastically populated. As a result, the correlation between \vphi{} and $\eta$ vanishes. 

As a companion to \autoref{fig:daily_mod}, \autoref{fig:eta_grid} shows $\tilde{\eta}$ for the highest-\vphi{} halo (\Subfind{} ID~455291, in the left column), the median-\vphi{} halo (ID~459558, in the center column), and the lowest-\vphi{} halo (ID~516101, in the right column). The Mollweide distributions are oriented such that $+\hat{\phi}$ is in the center, with dotted lines denoting the Galactic equator and every $90^\circ$ of latitude. As described in the main body, we convolve the distribution of $\eta$ with a kernel that emulates the angular resolution of an idealized directional DM detector, the result of which is denoted $\tilde{\eta}$. We also show the trajectory on the sky traced out by that detector's preferred axis: it follows a cone centered at the north celestial pole (i.e., aligned with Earth's rotation axis) that intersects the Earth's Galactocentric velocity vector (near $+\hat{\phi}$, marked with an asterisk in the figure). To guide the eye, we additionally show black markers every six hours along the trajectory. At low \vmin{}, in the top row, DM coming from many directions on the sky moves fast enough to cause a scattering event, leading to large $\eta$ but low on-sky variation. As \vmin{} increases, $\eta$ decreases, but the signal comes primarily from the direction of Earth's Galactocentric motion, increasing the on-sky variability. This increase is illustrated in \autoref{fig:fRMS}, which shows \fRMS{}, the root-mean-square variation of $\tilde{\eta}$ along the detector's trajectory, across a range of \vmin{}. Corotation generally suppresses \fRMS{}, as the DM signal comes from more directions on the sky than in an isotropic halo where the Earth's motion results in a collimated DM wind.

\phantomsection
\label{app:detector}
\section{Appendix B:\enspace{}Idealized Detector Response}
We adopt a simplified model for the angular response of a directional detector, in which we keep the leading-order anisotropic contribution in the multipole expansion of the material response (i.e., the squared form factor). For a scalar, spin-independent material response in the absence of time reversal--breaking fields, the dipole term is identically zero, with the leading anisotropic contribution in the quadrupole~\cite{Lillard:2026jdo}. We neglect the isotropic monopole term, thereby considering a maximally anisotropic benchmark in which the scattering rate is proportional to $\left|\hat{\mathbf{q}}\cdot\hat{\mathbf{n}}\right|^2$, for $\mathbf{q}$ the momentum transfer and $\hat{\mathbf{n}}$ the detector's preferred axis. Near kinematic threshold, energy conservation requires that the direction of momentum transfer be approximately aligned with the direction of the incident velocity, motivating our choice of point-spread function used in the text. It represents an idealized benchmark of the maximal modulation in a detector with a uniaxial response.

Such strongly anisotropic responses, which can present an $\mathcal{O}(1)$ fractional daily modulation near threshold, are expected in, for example, the planar electronic wave functions of aromatic molecular crystals~\cite{Blanco:2021hlm} and the anisotropic band structure of Dirac materials such as $\mathrm{ZrTe}_5$ and $\mathrm{Eu}_5\mathrm{In}_2\mathrm{Sb}_6$~\cite{Coskuner:2019odd,Hochberg:2017wce,Abbamonte:2025guf}. Additionally, low dimensional targets are also highly anisotropic. Quasi-2D targets like graphene and Quasi-1D targets such as aligned carbon nanotubes have well-studied, and strongly axis-selective detector responses~\cite{Cavoto:2019flp,Hochberg:2016ntt}. While these systems do not generically realize a purely quadrupolar response, they demonstrate that the maximally modulating limit considered here is a useful benchmark.
\end{document}